\documentstyle[prl,aps,epsfig]{revtex}
\tighten
\draft
\begin{document}
\twocolumn[\hsize\textwidth\columnwidth\hsize\csname
@twocolumnfalse\endcsname

\title{Hydrodynamic scaling from the dynamics of relativistic quantum field
theory}
\author{Lu\'{\i}s M. A. Bettencourt$^1$, Fred Cooper$^2$ and Karen Pao$^3$}
\address{$^1$Center for Theoretical Physics, Massachusetts Institute of
Technology, Bldg. 6-308, Cambridge MA 02139}
\address{$^2$Theoretical Division, MS B285, Los Alamos National Laboratory,
Los Alamos NM 87545}
\address{$^3$Applied Physics Division, MS F663, Los Alamos National
Laboratory, Los Alamos NM 87545}

\date{\today}

\maketitle

\begin{abstract}
Hydrodynamic behavior is a general feature of interacting systems with
many degrees of freedom constrained by conservation laws. To date
hydrodynamic scaling
in relativistic quantum systems has been observed in many high energy
settings, from cosmic ray detections to accelerators, with large
particle multiplicity final states. Here we show first evidence for the
emergence of hydrodynamic scaling in the dynamics of a relativistic
quantum field theory. We consider a simple scalar $\lambda \phi^4$ model in
1+1 dimensions in the Hartree approximation and study the dynamics of two
colliding kinks at relativistic speeds as well as the decay of a localized high energy
density region. The evolution of the energy-momentum tensor determines the
dynamical local equation of state and allows the measurement of the speed of sound.
Hydrodynamic scaling emerges at high local energy densities.
\end{abstract}

\pacs{PACS Numbers: 25.75.Ld, 05.70.Ln, 11.80.-m, 25.75.-q  \hfill
MIT--CTP-3179, LAUR-01-5183}

%
%
\vskip2pc]

Hydrodynamics has been used since the work of Landau \cite{Landau},
to describe the properties of high multiplicity final states in high energy
particle collisions. While Landau's motivations dealt with high energy
cosmic rays
there has been since ample evidence from accelerator experiments that
hydrodynamic
scaling (the longitudinal velocity $v_x = x/t$ ), which implies flat
rapidity
distributions, is the correct approximate kinematical constraint
for the dynamics of high energy particle collisions.

These findings imply that hydrodynamic scaling must emerge from the dynamics
of quantum field theories, if the latter are to be correct descriptions of
collective behavior in particle physics models. While the applicability of
quantum
field theory in these regimes is not in doubt, it has not been demonstrated
that
hydrodynamic scaling, which implies that that the energy density isosurfaces
are
surfaces of constant
$\tau^2 = (t^2-x^2)$, is achieved at sufficiently high center of mass
collision
energies.

Present and future experimental prospects for the study of hydrodynamic
scaling in high energy experiments are tremendous. The relativistic heavy
ion collider
(RHIC) is presently producing the highest energy, highest multiplicity
hadronic final states
ever accessible in a controlled environment \cite{RHICExperiment}.
The Large Hadron Collider (LHC) will later produce even more
spectacular events. The detailed understanding of hydrodynamic flows in
these experiments constitutes the most promising way for the determination
of the
thermodynamic properties of nuclear matter at high temperatures
\cite{RHICHydro}, {\it
viz.} its equation of state, and the nature of the confinement and chiral
symmetry
breaking transition.  The connection between  hydrodynamic scaling and flat
rapidity
distributions in the context of Landau's hydrodynamical model
was first discussed in \cite{ref:cfs}. In the context of ``boost
invariance",
the same scaling law was developed in detail by Bjorken \cite{ref:bjorken}.

Direct field theoretical methods, although still in their adolescence, offer
much promise for the understanding of hydrodynamic scaling and the limits of
its applicability. Moreover they make accessible regimes where particle
coherence is important, which escape Boltzmann particle methods.
In this letter we show, for the first time, how hydrodynamic scaling emerges
from the dynamics of a simple 1+1 dimensional scalar field theory in the
Hartree approximation. Our results allow us to map the equation of state as
a
function of space and time, and, under well known assumptions, determine the
speed of
sound $c_0$.

To exhibit the ubiquity of hydrodynamic scaling we study two different
situations: one in which a hot region is formed in the wake of the collision
of two leading particles (kinks) at relativistic velocities, and another
simpler one where we construct a local energy overdensity which is allowed
to relax
under its own self-consistent evolution.
To be definite we will be concerned with a scalar $\lambda \phi^4$ quantum
field theory with Lagrangian density
\begin{equation}
L= {1 \over 2} \partial_\mu \phi \partial^\mu \phi + {1 \over 2}\mu_\Lambda
^2
\phi^2 -{1 \over
4} \lambda \phi^4.
\label{eq:harlag}
\end{equation}
The well known Hartree approximation  is the simplest nontrivial truncation
of the coupled equations for the field's correlation functions, which
assumes that
all {\it connected} correlation functions beyond the second are negligible
\cite{Hartree}.
This leads to a dynamical equation for the mean field
$\varphi \equiv \langle \phi \rangle$ and the connected two-point function.
We write the quantum field $\phi = \varphi + \hat \psi$, where $\hat \psi$
are fluctuations, $\langle \hat \psi \rangle = 0$.
The equations of motion for $\varphi$ and the 2-point function
$G(x,y) =  \langle \hat \psi(x) \hat \psi(y)\rangle$ then are
\begin{eqnarray}
&&\left[ \Box  - \mu_\Lambda^2
+ {\lambda } \varphi^2(x) + 3 \lambda G(x,x) \right] \varphi(x)  =0, \\
&&\left[\Box  - \mu_\Lambda^2  + {3 \lambda } \left( \varphi^2(x) + G(x,x)
\right)
\right]G(x,y) = 0.
\nonumber
\end{eqnarray}
To solve the equation for the Green's functions we will rely on a complete
orthogonal mode basis $\psi_k(x)$
\begin{equation}
\hat \psi(x) = \sum_k \left[ a^\dagger_k \psi_k^*(x) +a_k \psi_k(x) \right],
\end{equation}
where $a_k^\dagger, a_k$ are creation and annihilation operators obeying
canonical commutation relations. In terms of the mode fields $\psi_k$,
at zero temperature
\begin{eqnarray}
G(x,y) = \sum_k ~\psi_k(x) \psi^*_k(y).
\end{eqnarray}

The effective mass squared of the propagator, $\chi(x,t)$  must be finite
which tells us how to choose the bare mass $\mu^2_\Lambda$.
In 1+1 dimensions the  self-energy has only  a logarithmic divergence, which
is eliminated by a simple mass renormalization. We choose
\begin{eqnarray}
 -  \mu_\Lambda^2 = \pm m^2 - 3 \lambda \int {dk \over 2 \pi}
{1 \over 2 \sqrt {k^2 + \chi}}\equiv  \pm m^2 -3 \lambda G_0,
\end{eqnarray}
leading to the existence of two homogeneous stable phases, corresponding to
$\chi_0=m^2$, $\phi=0$ and $\chi_0=2m^2$, $\phi^2=m^2/\lambda$, i.e. a
symmetric and broken symmetry phase, respectively. The renormalized
equations are
\begin{eqnarray}
&&\left[ \Box  + \chi(x) - 2 \lambda \varphi^2 (x) \right] \varphi(x)  = 0,
\nonumber \\
&&\left[ \Box  + \chi(x) \right] \psi_k(x,t)  = 0  \qquad \forall_k,
\label{eq:motion} \\
&&\chi(x) = \pm m^2 + 3 \lambda \varphi^2 (x) +3 \lambda G_R(x,x),
\label{eq:renorm}
\end{eqnarray}
where the renormalized $G_R(x,x)=G(x,x) - G_0$.
The challenge posed by
Eqs.~(\ref{eq:motion}-\ref{eq:renorm}) in spatially inhomogeneous cases is
that we need
to solve many partial differential  equations simultaneously. In a spatial
lattice of
linear size $L$, the computational  effort is of order $L^{2D}$, per time
step, where
$D$ is the number of space dimensions. Because of this demanding scaling we
focus on
$D=1$.

We consider two classes of initial
conditions: 1) Colliding kinks, in the broken phase where, at t=0 we have
\begin{eqnarray}
&& \varphi(x,t=0) = {m \over \sqrt{\lambda} } \varphi_{\rm kink}(x-x_0)
\varphi_{\rm kink}(-x-x_0), \\
&& \varphi_{\rm kink}(x) = \tanh \left( m x/\sqrt{2} \right),
\end{eqnarray}
with the kinks initially boosted towards each other at velocity $v$, and 2)
a Gaussian shape in the unbroken phase
\begin{eqnarray}
&& \varphi(x,0) = \varphi_0 \exp \left[ - {x^2 \over 2 A} \right], \quad
\partial_t \varphi(x,0) =0.
\label{eq:Gaussian}
\end{eqnarray}
In both cases we adopt at $t=0$ a Fourier plane-wave mode basis,
characteristic
of the unperturbed vacuum
\begin{eqnarray}
\psi_k (x,t) = {\sqrt{\hbar \over 2 \omega_k}}e^{i \left( k x + \omega_k t
\right)},
\qquad
\omega_k = \sqrt{k^2 + \chi_0}.
\end{eqnarray}
The orthonormality of the basis is preserved by the evolution,
Eqs.~(\ref{eq:motion}-\ref{eq:renorm})

To study the hydrodynamic behavior we need to specify the  operator energy
momentum tensor
$T^{\mu \nu}$
\begin{equation}
T^{\mu \nu} = \partial^{\mu} \phi \partial^{\nu} \phi - g^{\mu \nu}~L.
\end{equation}
Its expectation value, in terms of $\varphi$ and $\psi_k$, is
\begin{eqnarray}
&& \langle T_{00} \rangle=  {1\over 2} ({\varphi}_t) ^2 +{1\over 2}
({\varphi}_x) ^2
+ {1 \over 2} \sum_k\left[ |\psi_t^k|^2 + |\psi_x^k|^2 \right] + V_{\rm H},
\nonumber \\
&&\langle T_{11} \rangle =  {1\over 2} (\varphi_t) ^2 +{1\over 2}
(\varphi_x) ^2
+ {1 \over 2} \sum_k \left[ |\psi_t^k|^2 + |\psi_x^k|^2 \right] - V_{\rm H},
\nonumber \\
&& V_{\rm H} = { \chi^2 \over 12 \lambda} - \lambda {\varphi^4 \over 2},
\label{Tmunu} \\
&&\langle T_{01} \rangle = \langle T_{10} \rangle =
\varphi_t \varphi_x +{1 \over 2} \sum_k \left[\psi_x^k \psi^{\ast k}
_t+\psi_t^k
\psi_x^{\ast k} \right], \nonumber
\end{eqnarray}
where all arguments are at $x$. The subscripts $x,t$ are shorthand for
spatial and time derivatives, respectively. $T_{00}$ and $T_{11}$
contain two different types of  ultraviolet divergent contributions.
The  first arises from the 1-loop  integral $G(x,x)$.
 This logarithmic divergence is removed by mass renormalization
Eq.~(\ref{eq:renorm}). The second divergence appears in  the kinetic and
spatial
derivative fluctuation terms.  This divergence is purely quadratic, and is
already
present in the free field theory  in the vacuum sector.  By comparing the
mode sum with
a covariant dimensional  regularization scheme for the free field theory,
one deduces
that the correct subtraction  in the mode sum scheme is given by
\begin{equation}
{1 \over 2} \sum_k\left[ |\psi_t^k|^2 + |\psi_x^k|^2 \right]
\rightarrow {1 \over 2} \sum_k\left[ |\psi_t^k|^2 + |\psi_x^k|^2
- \vert k \vert \right]. \label{TmunuRenorm}
\end{equation}

In practice we discretize the fields $\varphi(x)$ and the set $\left\{
\psi_j \right\}$ on a spatial lattice with size $N$ and spacing $dx$ and
use periodic boundary conditions in space.
We choose $dx=0.125$, $N=1024$, and $m^2=1,\ \lambda=1, \ \hbar=1$.
The dynamical equations are solved using a symplectic fourth
order integrator (with a timestep $dt=0.025$).
With these choices, in the finite volume $L=Ndx$, the momentum $k$ 
takes a finite number of discrete values $k_n=\frac{2\pi n}{L}$, with 
$ n=\left\{-\frac{N}{2}, \ldots,
\frac{N}{2}-1 \right\}$ and continuum k-integrals become 
sums $\int dk/(2\pi) \to L^{-1}\sum_n$. The frequency $\omega_k$ now 
satisfies a lattice form of the dispersion relation, with 
\begin{equation}
\omega^2_k= \hat k^2+\chi^2, \;\;\;\;\hat k^2 = \frac{2}{dx^2}(1-\cos dx k_n).
\end{equation}
These forms also require that the renormalization of $T_{\mu
\nu}$ be achieved using appropriate lattice choices. In particular we adopt
$\vert k_n \vert = \sqrt{\hat k^2}$ in (\ref{TmunuRenorm}).

We are now ready to address the hydrodynamics of our field theory. Landau's
simplifying assumption, that pervades
hydrodynamic simulations of multi-particle flows in high energy experiments,
is that they behave collectively as a perfect fluid, corresponding to
\begin{eqnarray}
\langle T_{\mu \nu} \rangle =  (\varepsilon+ p) u^\mu  u^\nu - g^{\mu \nu}
p; ~~
\partial^\mu \langle T_{\mu \nu} \rangle = 0,
\label{idealfluid}
\end{eqnarray}
where  $u^\mu=\gamma(1,v)$, $(\gamma=1/\sqrt{1-v^2})$ 
is the collective fluid velocity and $\varepsilon$ and
$p$ are the comoving energy and pressure densities. The latter are
the eigenvalues of the energy momentum tensor, and can be obtained from the
invariance of its trace and determinant $\varepsilon - p = T^\mu _\mu$,
$\varepsilon p = {\rm Det}|T|$.
The fluid velocity can be obtained from the form (\ref{idealfluid})
\begin{equation}
T^{01} = (\varepsilon+ p) { v  \over 1-v^2}.
\end{equation}

A perfect fluid is the limiting hydrodynamic behavior
of a collisionless plasma, where transport coefficients, such as
viscosities are vanishing. In a mean-field approximation explicit collisions
are
neglected, so that the form  (\ref{idealfluid}) is compatible with our
dynamical
approximation.

\begin{figure}
\begin{center}
\begin{minipage}[t][7cm][t]{7cm}
\epsfig{figure=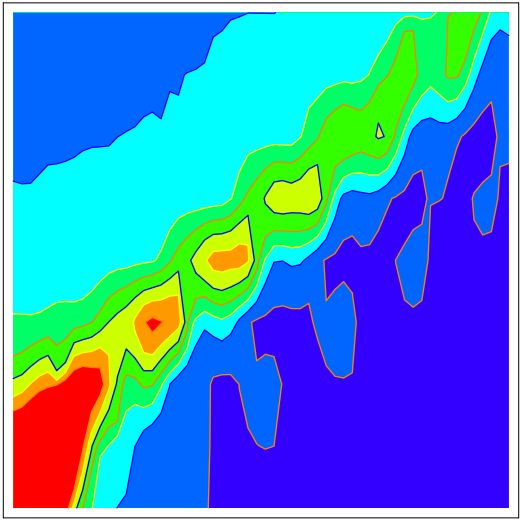,width=2.7in,height=2.7in}
\end{minipage}
\caption{Contours of equal energy density in space (horizontal) and time
(vertical),
near the collision  point of two kinks (at the origin of the plot),
initially boosted towards
each other at $v=0.8$. The collision is symmetric under spatial reflection.
We show the region
after and to the right of the collision point. Red denotes the highest
energy density,
navy blue the lowest.
The energy isosurfaces show signs of hydrodynamic scaling, following
approximate hyperboloids,
which are distorted because of the presence of the emerging kinks.}
\label{fig1}
\end{center}
\end{figure}

\begin{figure}
\begin{center}
\begin{minipage}[t][7cm][t]{7cm}
\epsfig{figure=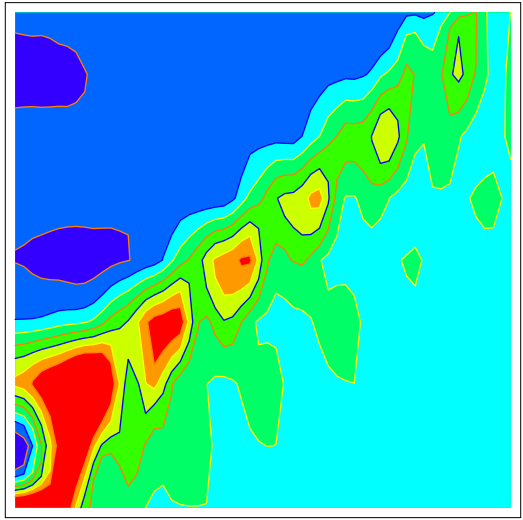,width=2.7in,height=2.7in}
\end{minipage}
\caption{Pressure isosurfaces in space-time for the situation
shown in Fig.\ref{fig1}. Asymptotically far from the kink trajectories
the equation of state mimics that of a gas with $p=c_0^2 \varepsilon$.
Close to the kinks the pressure gradients indicate space-time regions where
strong energy flows are imminent, such as the area around the kink collision
point,
at the origin of the plot.}
\label{fig2}
\end{center}
\end{figure}

\begin{figure}
\begin{center}
\begin{minipage}[t][7cm][t]{7cm}
\epsfig{figure=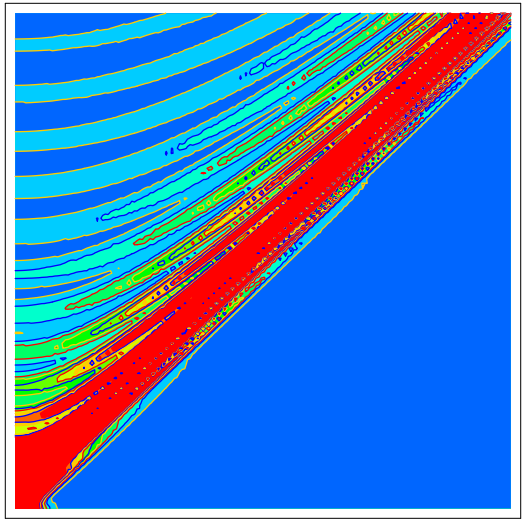,width=2.7in,height=2.7in}
\end{minipage}
\caption{$\varepsilon$ isosurfaces in space-time, for the decay
of an initial Gaussian shape (\ref{eq:Gaussian}), with $A=1$ and
$\varphi_0=5$.
The energy contained in the initial hot region is about two orders of
magnitude larger than that deposited by the kink collision, shown in
Figs.~\ref{fig1}-\ref{fig2}.
The red region carries away most of the energy in the form of a wave-packet
traveling
close to the speed of light. The energy isosurfaces follow exquisite
hyperboloids,
characteristic  of hydrodynamic scaling.}
\label{fig3}
\end{center}
\end{figure}

The attraction of scaling lies in the fact that the
relation $x=\pm vt$ allows for significant simplifications of the
hydrodynamic equations (\ref{idealfluid}), which can then be expressed in
terms
of a single variable and thus become ordinary differential equations.
These can then be solved analytically \cite{ref:cfs}, generating predictions
for the spatio-temporal behavior of hydrodynamic quantities such as energy
density $\varepsilon$, temperature or entropy. For example, for equations of
state
where $dp/d\varepsilon = c_0^2$, where $c_0$ is the (constant) speed of
sound, one easily
derives in 1D \cite{ref:cfs}
\begin{equation}
\varepsilon(x,t)/ \varepsilon_0 =\left( {\tau / x_0} \right)^{-
(1+c_0^2)};~~ \tau=\sqrt{t^2-x^2},
\label{Escaling}
\end{equation}
where $\varepsilon_0, x_0$ are integration constants.
The equation of state $p=c_0^2 \varepsilon$ can be obtained using simple
assumptions
about the (hadronic) excitation spectrum \cite{ref:BU,ref:Shuryak}.
Eq.~(\ref{Escaling}) implies that energy density isosurfaces lie on
hyperboloids $t^2-x^2 =$const, a property that we can easily check in our
results, see
Figs.~\ref{fig1} and \ref{fig3} corresponding to initial conditions 1) and
2), respectively.
Fig.~\ref{fig2} shows the pressure in space-time for the same situation as
in Fig.~\ref{fig1}.

Besides the shape of energy isosurfaces hydrodynamic scaling also predicts
that the single particle distribution functions are flat in the particle rapidity 
variable
$y=\frac{1}{2}\ln[(E+p_\parallel)/(E-p_\parallel)]$, where $E$ and
$p_\parallel$ refer
to the energy and momentum in the direction of the collision of 
an outgoing particle. This result is seen  experimentally  in the central 
rapidity region \cite{flatrapidity}, and can be interpreted as arising 
either from the approximate boost invariance of two highly Lorentz contracted 
colliding nuclei at high energies  \cite{ref:bjorken}, or from the fact that, 
in the center of mass frame, the Lorentz contracted source for particle production
has a  negligible longitudinal size, when compared to the asymptotic particle's
(pion) Compton wavelength \cite{Landau,ref:cfs}. 
At sufficiently high deposited energies the two approaches lead to the same results. 

Scaling solutions do not preserve global energy conservation. 
Thus energy isosurfaces must
eventually deviate from the scaling hyperbola and join neighboring
isosurfaces,
creating characteristic horn like shapes. This behavior, which can be
extracted directly
from hydrodynamic equations, is also observed in the quantum field
solutions, see
Figs.~\ref{fig1} and \ref{fig3}.

\begin{figure}
\begin{center}
\begin{minipage}[t][8cm][t]{9cm}
\epsfig{figure=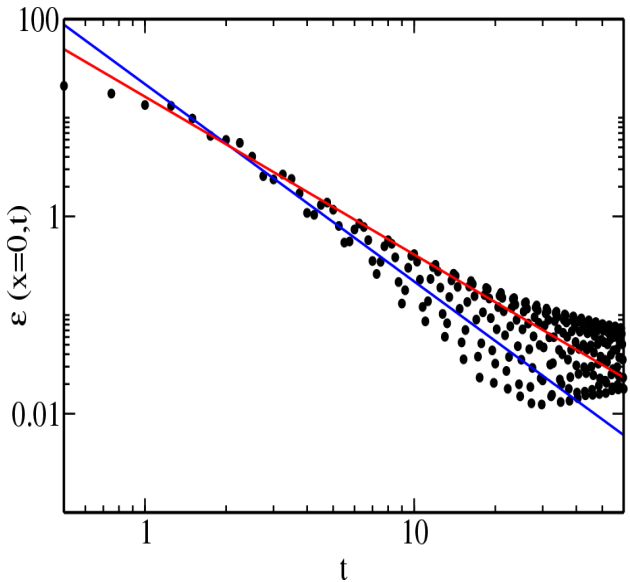,width=3.in,height=3.25in,angle=270}
\end{minipage}
\caption{$\varepsilon(x=0,t)$, in the wake of the decay of a
Gaussian hot region of Fig.~\ref{fig3}. The solid lines show power laws of
the form
(\ref{Escaling}) with $1+c_0^2=2$ (blue), corresponding to the
ultrarelativistic ideal gas
($c_0=1$) and with $1+c_0^2=1.6$. }
\label{fig4}
\end{center}
\end{figure}

Eq.~(\ref{Escaling}) also allows us a measurement of the speed of sound
$c_0$. Fig.~\ref{fig4} shows the decay in time of the energy density of the
Gaussian mean-field
profile already discussed in the context of Fig.~\ref{fig3}.
The resulting fit suggests a value of $1\geq c_0 \stackrel{>}{{}_\sim}
0.77$, compatible with an
ultrarelativistic equation of state $c_0=1$ in 1D.
In conclusion we have demonstrated for the first time, in a variety of
settings, that hydrodynamic scaling emerges from the dynamics of quantum
field
theory at sufficiently high energy densities.
We analyzed situations both with leading particles, which constitute the
asymptotic states both before and after the collision, and following the
evolution
of a simple local energy overdensity.
The extension of this type of calculation to 3D, where hydrodynamics is
richer,
and to include scattering, necessary for the description of real
fluids, remain necessary steps to make real time studies of quantum fields
predictive
experimentally in the context of heavy ion collisions.
Nevertheless studying the real time dynamics of quantum field theories
demonstrates
the applicability of models of particle physics in the largely unexplored
limits of very large
time and spatial scales, where they acquire fascinating macroscopic and
non-perturbative
properties.

We thank K.~Rajagopal for comments on the manuscript.
This work was supported in part by the D.O.E. under research agreement
$\#$DF-FC02-94ER40818. Numerical work was done at the Nirvana cluster at
LANL.

\end{document}